\documentclass[twocolumn,PRB,aps,psfig,preprintnumbers,superscriptaddress]{revtex4}
\usepackage{graphicx}
\usepackage{epstopdf}
\usepackage{float}
\usepackage{color}
\usepackage{amsmath}

\begin{document}

\title{Inhomogeneous magnetic ordered state and evolution of magnetic fluctuations in Sr(Co$_{1-x}$Ni$_x$)$_2$P$_2$ revealed by $^{31}$P NMR}

\author{Nao Furukawa}
\affiliation{Ames National Laboratory, U.S. DOE, Ames, Iowa 50011, USA}
\affiliation{Department of Physics and Astronomy, Iowa State University, Ames, Iowa 50011, USA}

\author{Qing-Ping Ding}
\affiliation{Ames National Laboratory, U.S. DOE, Ames, Iowa 50011, USA}

\author{Juan Schmidt}
\affiliation{Ames National Laboratory, U.S. DOE, Ames, Iowa 50011, USA}
\affiliation{Department of Physics and Astronomy, Iowa State University, Ames, Iowa 50011, USA}

\author{Sergey L. Bud'ko}
\affiliation{Ames National Laboratory, U.S. DOE, Ames, Iowa 50011, USA}
\affiliation{Department of Physics and Astronomy, Iowa State University, Ames, Iowa 50011, USA}

\author{Paul C. Canfield}
\affiliation{Ames National Laboratory, U.S. DOE, Ames, Iowa 50011, USA}
\affiliation{Department of Physics and Astronomy, Iowa State University, Ames, Iowa 50011, USA}

\author{Yuji Furukawa}
\affiliation{Ames National Laboratory, U.S. DOE, Ames, Iowa 50011, USA}
\affiliation{Department of Physics and Astronomy, Iowa State University, Ames, Iowa 50011, USA}

\date{\today}

\begin{abstract}
 
SrCo$_2$P$_2$ with a tetragonal structure is known to be a Stoner-enhanced Pauli paramagnetic metal being nearly ferromagnetic. 
      Recently Schmidt  {\it et~al.} [Phys. Rev. B {\bf 108}, 174415 (2023)] reported that a ferromagnetic ordered state is actually induced by a small Ni substitution for Co of $x$  = 0.02 in Sr(Co$_{1-x}$Ni$_x$)$_2$P$_2$ where antiferromagnetic ordered phase also appears by further Ni-substitution with $x = 0.06-0.35$.   
      Here, using nuclear magnetic resonance (NMR) measurements on  $^{31}$P nuclei, we have investigated how the magnetic properties change by the Ni substitution in  Sr(Co$_{1-x}$Ni$_x$)$_2$P$_2$ from a microscopic point of view, especially focusing on the evolution of magnetic fluctuations with the Ni substitution and the characterization of the magnetically ordered states.
   The temperature dependences of  $^{31}$P spin-lattice relaxation rate divided by temperature ($1/T_1T$) and Knight shift ($K$) for SrCo$_2$P$_2$ are reasonably explained by a model where a double-peak structure for the density  of states near the Fermi energy is assumed.
     Based on a Korringa ratio analysis using the $T_1$ and  $K$  data,  ferromagnetic spin fluctuations are found to dominate in the ferromagnetic Sr(Co$_{1-x}$Ni$_x$)$_2$P$_2$  as well as the antiferromagnets where no clear antiferromagnetic fluctuations are observed. 
    We also found  the distribution of the ordered Co moments in the magnetically ordered states from the analysis of  the $^{31}$P NMR spectra exhibiting a characteristic rectangular-like shape.

\end{abstract}

\maketitle

\section{Introduction}

    Magnetic fluctuation is one of the important parameters to control the physical properties of unconventional superconductors (SCs) such as high $T_{\rm c}$ cuprates and iron-based SCs.
     The discovery of superconductivity in iron pnictides in 2008 \cite{Kamihara2008} triggered considerable experimental and theoretical attention  on  transition-metal (TM) pnictides  \cite{Johnston2010,Canfield2010,Stewart2011}. 
   Among them, Co compounds with ThCr$_2$Si$_2$-type structure, $A$Co$_2$X$_2$ ($A$ = alkaline earth or lanthanide, $X$ = pnictogen)  
have been found to show a rich variety of magnetic properties with different crystal structures.
    SrCo$_2$P$_2$ with an uncollapsed tetragonal  (ucT) structure shows no magnetic ordering and is a Stoner-enhanced Pauli paramagnet with dominant ferromagnetic interactions \cite{Moresen1998, Jia2009, Imai2014, Imai2015PP, Imai2017}.
    In contrast, an A-type antiferromagnetic (AFM) state has been reported in CaCo$_2$P$_2$ with a collapsed tetragonal (cT) structure below a N\'eel temperature of 110 K, in which the Co moments are ferromagnetically aligned in the $ab$ plane and the moments adjacent along the $c$ axis are aligned antiferromagnetically \cite{Reehuis1998,Imai2017,Baumbach2014PRB}. The ordered Co moments in the AFM state are reported  to be 0.32-0.35 $\mu_{\rm B}$ from the neutron diffraction (ND) measurements \cite{Reehuis1998} and nuclear magnetic resonance (NMR) measurements \cite{Higa2018}.
   A ferromagnetic (FM) state is also reported  by replacing the divalent alkaline-earth ions by the trivalent La ions in the metallic LaCo$_2$P$_2$ with an ucT structure below  a Curie temperature of $T_{\rm C}$ = 130 K where the saturated Co moment is reported to be of 0.4 $\mu_{\rm B}$ \cite{Reehuis1994,Imai2015PRB, Teruya_LaCo2P2}.

    Magnetic ordering in Co-based compounds was also reported to be induced by Ca substitution for Sr in (Sr$_{1-x}$Ca$_x$)Co$_2$P$_2$ \cite{Jia2009, Imai2017, Sugiyama2015PRB}
 and also by Ge substitution for P in SrCo$_2$(P$_{1-x}$Ge$_x$)$_2$ \cite{Moriyama2023}.
    In the case of (Sr$_{1-x}$Ca$_x$)Co$_2$P$_2$, although  the Stoner-enhanced Pauli paramagnetic state is observed in the ucT phase with $x = 0-0.5$, the A-type AFM ordering appears in the cT phase with $x = 0.6-1$ \cite{Jia2009}, indicating a strong relationship between the magnetism and crystal structure. 
     It is interesting to point out that, in contrast to the Co compounds, no magnetic order  has been observed in the cT  phase in Fe-based compounds such as CaFe$_2$As$_2$ \cite{Kreyssig2008,Pratt2009,Prokes2010,Ran2011} where AFM fluctuations are also revealed to be completely suppressed \cite{Kawasaki2010,Kawasaki2011,Soh2013,Furukawa2014}. 
    In the case of SrCo$_2$(P$_{1-x}$Ge$_x$)$_2$, a ferromagnetic ordered state has been observed in $x$ = 0.42-0.61 which is an intermediate  region between the ucT (the SrCo$_2$P$_2$ side)  and the cT (the SrCo$_2$Ge$_2$ side) phases  and a ferromagnetic quantum criticality  has been discussed \cite{Moriyama2023}.



\begin{figure}[b]
\includegraphics[width=8.5cm]{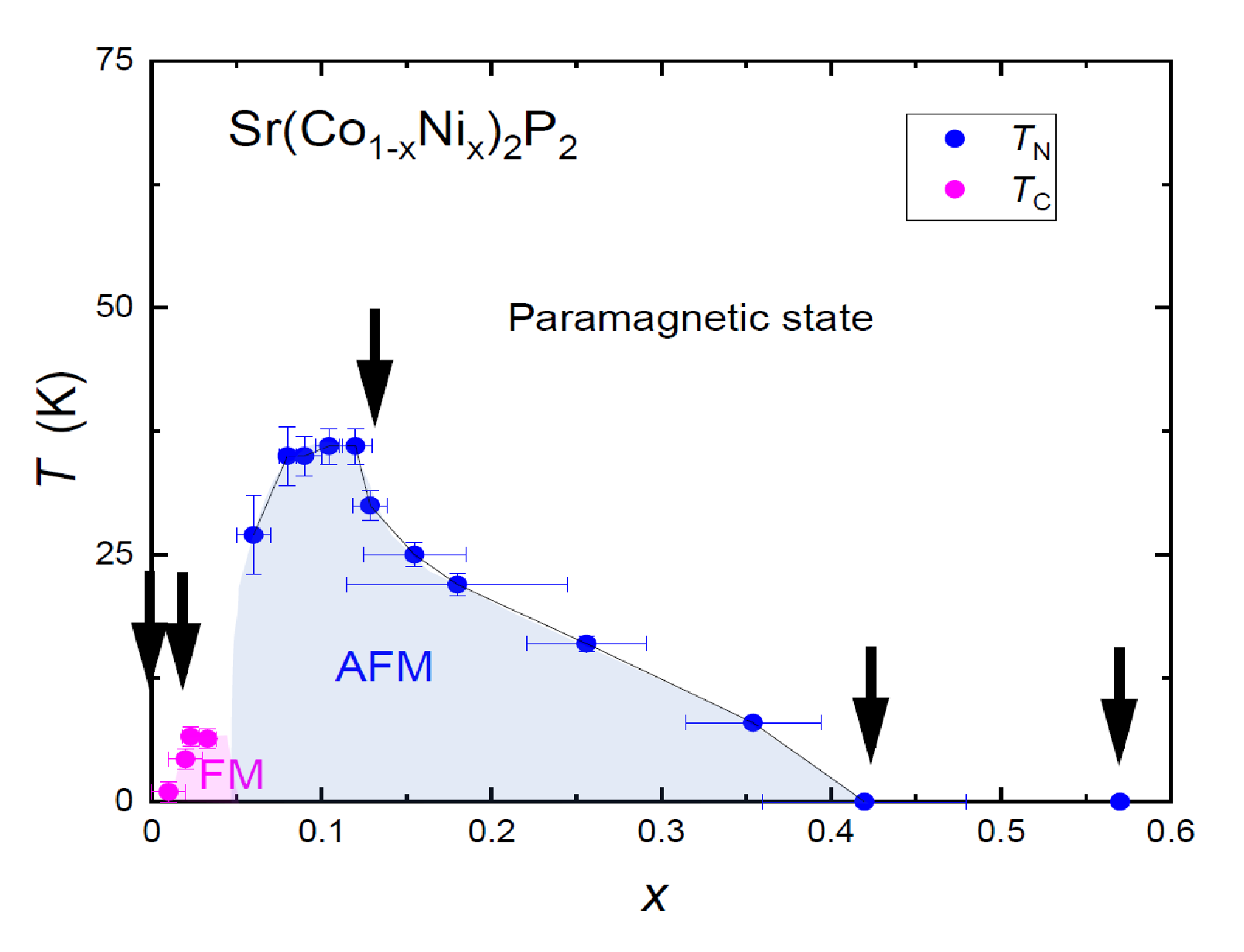} 
\caption{Phase diagram of Sr(Co$_{1-x}$Ni$_x$)$_2$P$_2$. $T_{\rm N}$ and $T_{\rm C}$ are from Ref. \cite{Schmidt2023}. 
AFM and FM represent the antiferromagnetic and ferromagnetic ordered states at a low magnetic field of 100 Oe, respectively.  
   Arrows indicate the positions of the representative of the Ni substitution levels used in the present work.}
\label{fig:PhaseDiagram}
\end{figure}

    Recently it was shown that, from the detailed investigation of the Ni substituted SrCo$_2$P$_2$ compounds,  Sr(Co$_{1-x}$Ni$_x$)$_2$P$_2$ can be systematically tuned to have one of three different magnetic ground states: AFM, FM and  paramagnetic states by electron doping \cite{Schmidt2023}.
      Figure 1 shows a schematic phase diagram of Sr(Co$_{1-x}$Ni$_x$)$_2$P$_2$ determined by magnetization and resistivity measurements \cite{Schmidt2023}.
      Although $x$ = 1 (SrNi$_2$P$_2$) exhibits a one-third collapsed orthorhombic  structure  (tcO) \cite{Schmidt2023}, 
the tcO phase is rapidly suppressed with Co additions and the magnetically ordered phases appear only in the ucT phase, which is different from the aforementioned other systems (Sr$_{1-x}$Ca$_x$)Co$_2$P$_2$  and  SrCo$_2$(P$_{1-x}$Ge$_x$)$_2$. 
      With only a 2 \% Ni substitution to the nonmagnetic SrCo$_2$P$_2$, ordering with a ferromagnetic component is induced with  $T_{\rm C}$ = 4.3-6.6 K.   
      With further Ni substitution, the AFM state appears for  0.06 $\leq$ $x$ $\leq$  0.35 with the highest $T_{\rm N}$ = 36 K around $x$ = 0.1-0.12. 
     As the magnetic properties of Co compounds seem to largely depend on crystal structure as well as the valence number of Co ions,   Sr(Co$_{1-x}$Ni$_x$)$_2$P$_2$ is a good system to investigate only the carrier doping effect on the magnetic properties without changing the ucT crystal structure. 
      Up to now, no investigation of the magnetic properties from a microscopic point of view or magnetic fluctuations in  Sr(Co$_{1-x}$Ni$_x$)$_2$P$_2$ has been reported. 
     Since magnetic fluctuations are considered to be one of the key ingredients to characterize the physical properties of materials, it is crucial to reveal how magnetic fluctuations  vary with Ni substitution  in Sr(Co$_{1-x}$Ni$_x$)$_2$P$_2$.
    NMR is an ideal tool for the microscopic study of the magnetic properties and the low-energy magnetic fluctuations in correlated electron systems. 
     It is known that the nuclear spin-lattice relaxation rate (1/$T_1$) reflects the wave vector $q$-summed dynamical susceptibility. 
   On the other hand, NMR spectrum measurements, particularly  the Knight shift ($K$), give us information on the static magnetic susceptibility $\chi$. 
   Thus from the temperature dependence of 1/$T_1$ and $K$, one can obtain valuable insights into magnetic fluctuations in materials. 

      In this paper, we carried out  $^{31}$P NMR measurements to examine the local magnetic  properties of Sr(Co$_{1-x}$Ni$_x$)$_2$P$_2$. 
  The characteristic  temperature dependences of  $K$  and 1/$T_1T$ exhibiting a double-maxima around 25 K and 125 K in the parent compound SrCo$_2$P$_2$   were reasonably explained by a model where a double-peak structure of the density of states (DOS) near the Fermi energy is assumed. 
  Our analysis, based on the modified Korringa relation, reveals  electron correlations enhanced  around FM wavenumber $q$ = 0  in SrCo$_2$P$_2$. 
   We find the FM spin fluctuations are strongly enhanced with the only 2 \% Ni substitution in the ferromagnetic Sr(Co$_{0.98}$Ni$_{0.02}$)$_2$P$_2$. 
   Similar strong FM spin fluctuations are also found to dominate in an antiferromagnet Sr(Co$_{0.87}$Ni$_{0.13}$)$_2$P$_2$ with $T_{\rm N}$ = 30 K.  
  With further Ni substitution up to $x$ = 0.57 through $x$ = 0.42, such FM spin fluctuations as well as the magnetic  ordering completely disappear, leading to a noncorrelated metal where a typical Korringa behavior of 1/$T_1T$ = const. was observed without showing any indication of quantum critical behavior around $x$ = 0.42 in Sr(Co$_{1-x}$Ni$_x$)$_2$P$_2$.
    Before proceeding, we would like to point out that in Ref. \cite{Schmidt2023} samples were presented as Sr(Ni$_{1-x}$Co$_x$)$_2$P$_2$; in this work we present our data based on the formula Sr(Co$_{1-x}$Ni$_x$)$_2$P$_2$ since the magnetic ordering we are interested in develops for $x$ as small as 0.02 and it disappears  before $x$  = 0.50.

\section{Experimental details}
      The single crystals of Sr(Co$_{1-x}$Ni$_x$)$_2$P$_2$  ($x$ = 0, 0.02, 0.13, 0.42 and 0.57) used in this study were grown out of Sn flux \cite{Schmidt2023}, using conventional high-temperature growth techniques \cite{Canfield2020}. 
    The sizes of the crystals are similar ($\sim$$3\times4\times0.1$ mm$^3$).
   The Ni substitution levels of the single crystals used in this study were determined by wavelength dispersive x-ray spectroscopy, and the crystals are characterized by magnetization and resistivity measurements \cite{Schmidt2023}. 
   No magnetic ordering was observed for $x$ = 0 (pure SrCo$_2$P$_2$), $x$ = 0.42  or 0.57 for $T$ $>$  2 K from the magnetization measurements, while a ferromagnetic ordering at $T_{\rm C}$ = 4.3  K and an antiferromagnetic ordering at $T_{\rm N}$ = 30 K were reported under a weak magnetic field of 100 Oe for our crystals $x$ = 0.02 and 0.13, respectively \cite{Schmidt2023}.   
   Details of the characterization of the crystals are  reported in Ref. \cite{Schmidt2023}.

   $^{31}$P (nuclear spin $I$ = 1/2 and gyromagnetic ratio $\frac{\gamma_{\rm N}}{2\pi}$ = 17.237 MHz/T)  NMR spectra were obtained either by fast Fourier transform (FFT) of the NMR echo signals or by plotting spin-echo intensity with changing resonance frequency under an external magnetic field  $H$ of  7.4089 Tesla.
 When the $^{31}$P NMR spectra become very broad due to magnetic ordering for $x$ = 0.02 and 0.13, we measured  the $^{31}$P NMR spectra  by sweeping the external magnetic field  $H$ at a constant resonance frequency.
   The NMR spectra for $x$ = 0.42 were also measured by sweeping the $H$.   
The magnetic field was applied parallel to either the crystal $c$ axis or the $ab$ plane, and the direction of the magnetic field in the $ab$ plane was not controlled.
   The origin of the Knight shift, $K$ = 0, of the $^{31}$P  nucleus was determined by the $^{31}$P  NMR measurements of H$_3$PO$_4$. 

  The $^{31}$P spin-lattice relaxation rate $(1/T_{1})$ was measured  at the peak position of the the spectra with a saturation recovery method.
   $1/T_1$ at each temperature ($T$) was determined by fitting the nuclear magnetization $M_{\rm N}$ versus time $t$  using the exponential function $1-M_{\rm N}(t)/M_{\rm N}(\infty) = e^ {-t/T_{1}}$,  where $M_{\rm N}(t)$ and $M_{\rm N}(\infty)$ are the nuclear magnetization at time $t$ after the saturation and the equilibrium nuclear magnetization at $t$ $\rightarrow$ $\infty$, respectively. 
   All the nuclear magnetization recovery curves  measured for all crystals of Sr(Co$_{1-x}$Ni$_x$)$_2$P$_2$  ($x$ = 0, 0.02, 0.13, 0.42 and 0.57) were well fitted with the function as shown in the supplementary materials \cite{SM}. 

\begin{figure}[h!tb]
\includegraphics[width=\columnwidth]{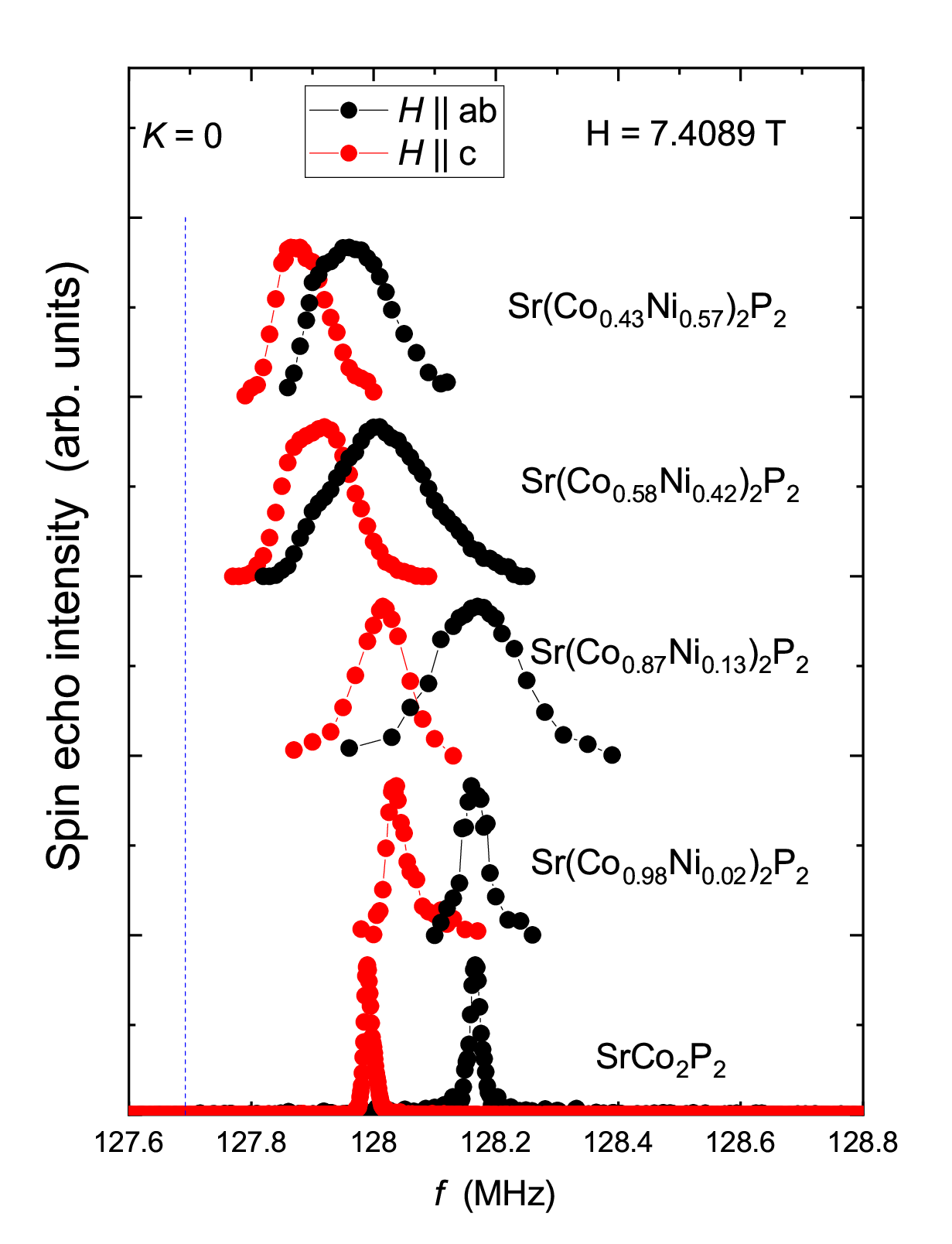}
\caption{Frequency-swept $^{31}$P-NMR spectra for magnetic fields $H\parallel c$ axis (red) and $H\parallel ab$ plane (black) in Sr(Co$_{1-x}$Ni$_x$)$_2$P$_2$ measured near room temperature (300 K for $x$ = 0 and 0.13, 290 K for $x$ = 0.02, 0.42 and 0.57).  The vertical dashed line represents the zero-shift position ($K=0$).}
\label{fig:spectrumRT}
\end{figure}

\begin{figure*}[h!tb]
\includegraphics[width=2\columnwidth]{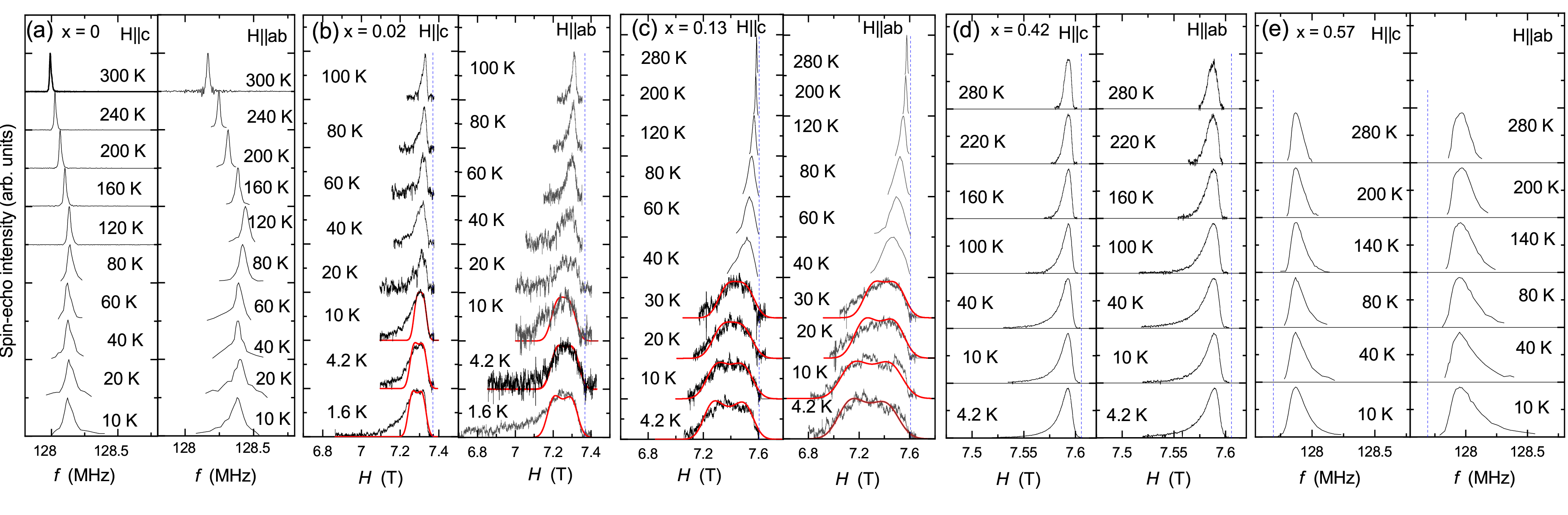}
\caption{Temperature dependence of $^{31}$P-NMR spectra under the magnetic fields $H\parallel c$ and $H\parallel ab$  in Sr(Co$_{1-x}$Ni$_x$)$_2$P$_2$. (a) $x$ = 0, (b) $x$ = 0.02, (c) $x$ = 0.13, (d) $x$ = 0.42  and (e) $x$ = 0.57.
 The blue vertical dashed lines in (b), (c), (d) and (e) represents the corresponding zero-shift positions in the $^{31}$P NMR spectra ($K=0$). 
For (a) and (e), the frequency-swept spectra were measured under the constant magnetic field of $H$ = 7.4089 T. The field-swept spectra  were measured with constant frequency of 127 MHz for (b)  and  131.1 MHz for (c) and (d), respectively.  For (c), the spectra above 40 K are replotted from the frequency-swept NMR spectra at $H$ = 7.4089 T where the horizontal axis for the spectra was changed from frequency to magnetic field by using the $\gamma_{\rm N}$ value of P nucleus. The red curves in (b) and (c) are simulated spectra (see the text for details).}
\label{fig:T-dep_spectrum}
\end{figure*}

\section{Experimental results}
\subsection{$^{31}$P NMR spectrum}
      Figure \ref{fig:spectrumRT} shows the frequency-swept $^{31}$P-NMR spectra of Sr(Co$_{1-x}$Ni$_x$)$_2$P$_2$  at near room temperature  for magnetic fields parallel to the $c$~axis ($H\parallel c$) and the $ab$~plane ($H\parallel ab$).  
     Since $^{31}$P nucleus has  $I$ = 1/2, a single line was observed as shown in the figure. 
     In the case of SrCo$_2$P$_2$ ($x$ = 0),  relatively sharp lines are observed with the full-width at half maximum (FWHM) values of  13.6 kHz and 19.6 kHz for $H\parallel c$ and $H\parallel ab$, respectively,  at room temperature.   
     With the Ni substitution for Co in Sr(Co$_{1-x}$Ni$_x$)$_2$P$_2$, the observed spectra become broader to 34.0 kHz and 36.0 kHz for $x$ = 0.02, 92.0 kHz and 154 kHz for $x$ = 0.13, 132 kHz and 189  kHz for $x$ = 0.42 and 99.0 kHz and 150 kHz for $x$ = 0.57, for $H\parallel c$ and $H\parallel ab$, respectively.
     This is mainly due to a distribution of the hyperfine coupling constant produced by the Ni substitution and also a distribution of the Ni concentration. 
     It is noted that the values of the FWHM for $x$ = 0.42 are slightly greater than those of others. 
    This originates from a slightly larger distribution of Ni concentration for $x$ = 0.42 crystals than others reported in Ref. \cite{Schmidt2023}.   
      It is worth mentioning that the NMR spectra  indicate no obvious phase separation in our crystals, as we do not observe any separated lines in the spectra. 

    The typical temperature dependence of $^{31}$P NMR spectra in Sr(Co$_{1-x}$Ni$_x$)$_2$P$_2$ is shown in Fig. \ref{fig:T-dep_spectrum}. 
    For $x$ = 0 [Fig. \ref{fig:T-dep_spectrum}(a)], although the NMR  lines are relatively sharp at high temperatures, the spectra become slightly broader with decreasing temperature  due to an inhomogeneous magnetic broadening.
This probably originates from a slight inhomogeneity of the hyperfine coupling constant due to some defects. 
     A similar inhomogeneous broadening was observed in $^{59}$Co NMR spectra of the isostructural compound SrCo$_2$As$_2$ \cite{Wiecki2015}.    
    For the magnetically ordered compounds with $x$ = 0.02 and 0.13, the observed spectra become very broad at low temperatures due to the magnetic ordering as shown in Figs. \ref{fig:T-dep_spectrum}(b)  and \ref{fig:T-dep_spectrum}(c), evidencing the the magnetic ordering in both samples from a microscopic point of view.
    We also note that the spectra become  asymmetric. 
   Since $^{31}$P nucleus has  $I$ = 1/2 and we used single crystals for our measurements, the asymmetric shape indicates a distribution of the hyperfine coupling constants at the P sites.
     It is interesting to point out that the shape of the NMR line in the magnetically ordered state for $x$ = 0.13 shows a characteristic rectangular-like shape [for example, see the spectra at the bottom in Fig. \ref{fig:T-dep_spectrum}(c)]. 
   Such a rectangular shape of the NMR spectrum is expected for powder samples of antiferromagnets \cite{Yamada1986,Devi2022}. 
 However, as we used singe crystals for our measurements, this cannot be invoked to explain the NMR line shape observed. 
It is also noted that, although the plateau region is less than that of $x$ = 0.13, a similar rectangular-like shape is observed in $x$ = 0.02 at low temperatures. 
We will discuss the shape of the NMR line below.      
    For the non-magnetic compound with $x$  = 0.42 and 0.58, although the observed spectra are asymmetric due to the  distribution of hyperfine coupling and Ni contents, no obvious change in the peak positions was observed, consistent with the non-magnetic ground state reported from the magnetization measurements \cite{Schmidt2023}.

 \begin{figure}[h!tb]
\centering
\includegraphics[width=\columnwidth]{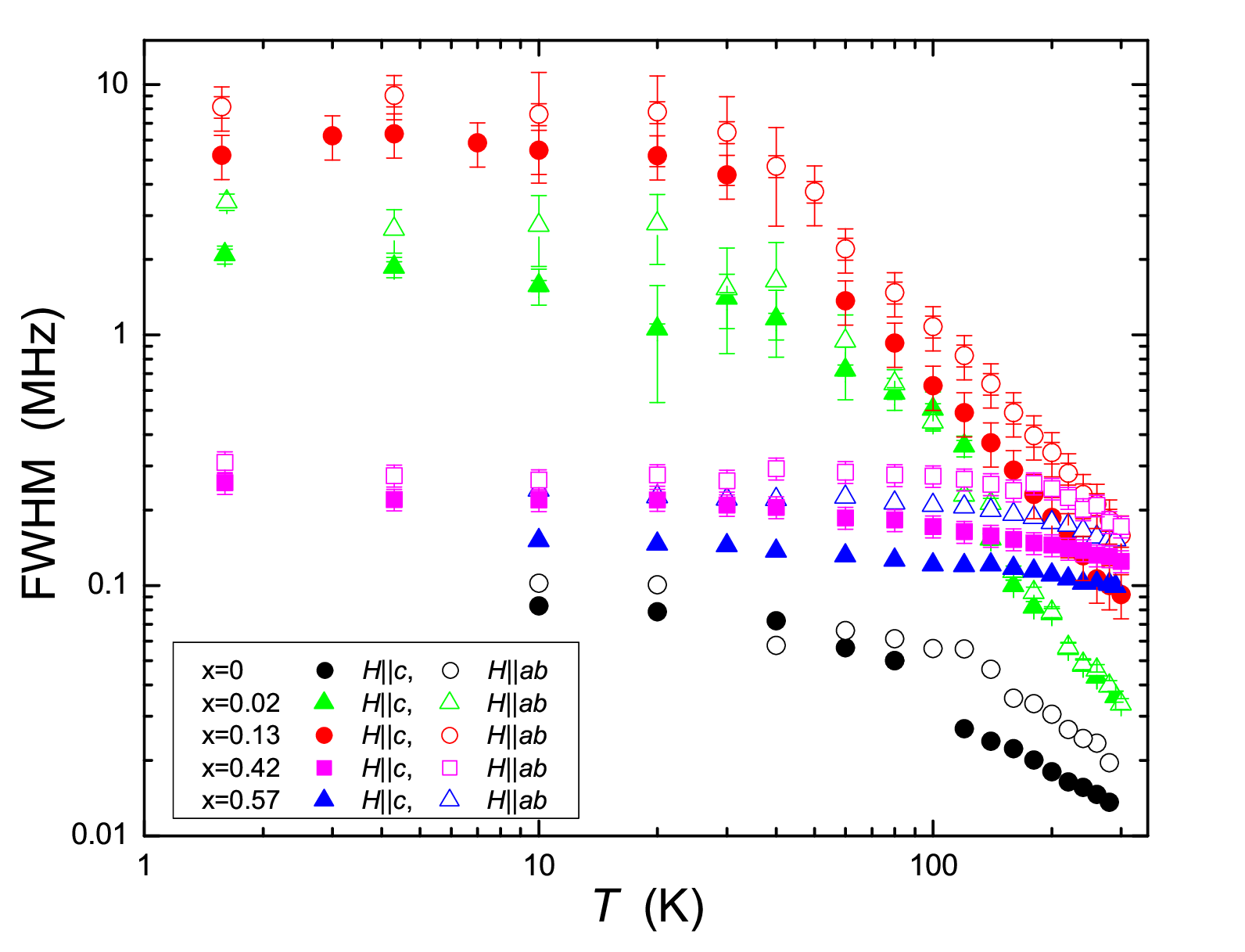}
\caption{Temperature dependence of full width at half maximum (FWHM) of $^{31}$P-NMR spectra under the magnetic field parallel to the $c$ axis ($H\parallel c$) and to the $ab$ plane ($H\parallel ab$)  in Sr(Co$_{1-x}$Ni$_x$)$_2$P$_2$.}
\label{fig:FWHM}
\end{figure}

Figure \ref{fig:FWHM} shows the temperature dependence of FWHM of $^{31}$P-NMR spectra under both magnetic field directions  $H\parallel c$ and $H\parallel ab$  for  Sr(Co$_{1-x}$Ni$_x$)$_2$P$_2$. 
   The values of FWHM of the field-swept NMR spectra for $x$ = 0.02, 0.13 and 0.42 were converted from Tesla to MHz by using the relation $f$ = $\frac{\gamma_{\rm N}}{2\pi}$$H$. 
  As described above, in the case of $x$ = 0.13,  the values of FWHM for both $H$ directions increase with decreasing temperature and level off below $T_{\rm N}$ = 30 K due to the magnetic ordering.  
  Interestingly similar temperature dependence of FWHM, but with different magnitudes, is observed in $x$ = 0.02. 
  This suggests that the magnetically ordered state at $x$ = 0.02 is well developed below $\sim$ 10 K under a magnetic field of $\sim$ 7.4 T, although the compound is reported to show  a ferromagnetic ordered state below $T_{\rm C}$ = 4.3 K under $H$ = 0.01 T.      

 \begin{figure}[h!tb]
\centering
\includegraphics[width=\columnwidth]{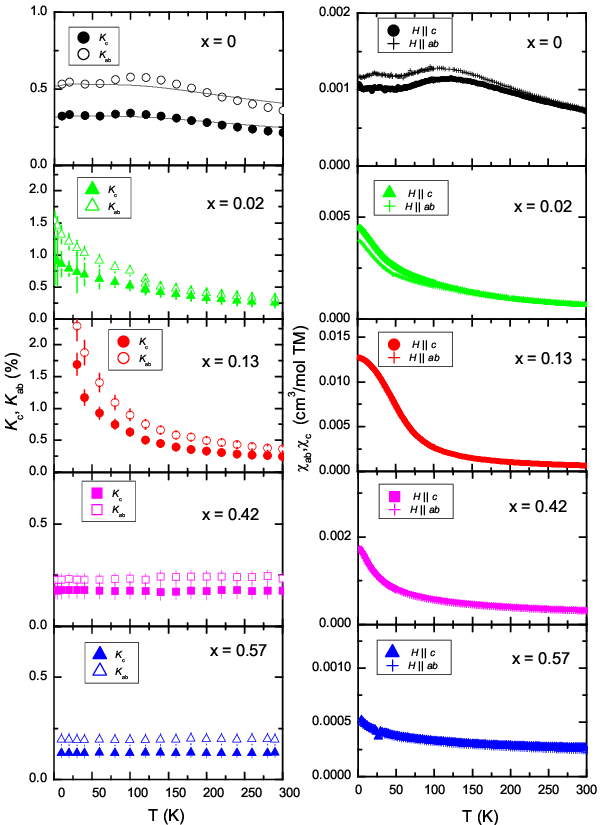}
\caption{Left panels: Temperature dependence of Knight shift for magnetic fields $H\parallel c$ axis ($K_c$)  and $H\parallel ab$ plane ($K_{ab}$) for Sr(Co$_{1-x}$Ni$_x$)$_2$P$_2$.  The black curves are the calculated  temperature dependence of  Knight shift assuming a double-peak structure in the density of states  near Fermi energy (see the text for details). 
Right panels: Temperature dependence of  magnetic susceptibility in units of cm$^3$/mol per transition metal (TM) ions  under a magnetic field of 7 T for  $H\parallel c$ axis ($\chi_c$)  and $H\parallel ab$ plane ($\chi_{ab}$) in Sr(Co$_{1-x}$Ni$_x$)$_2$P$_2$ except for the $\chi$ data for $x$ = 0.57 measured at 1 T. }
\label{fig:K}
\end{figure}

 \begin{figure}[h!tb]
\centering
\includegraphics[width=\columnwidth]{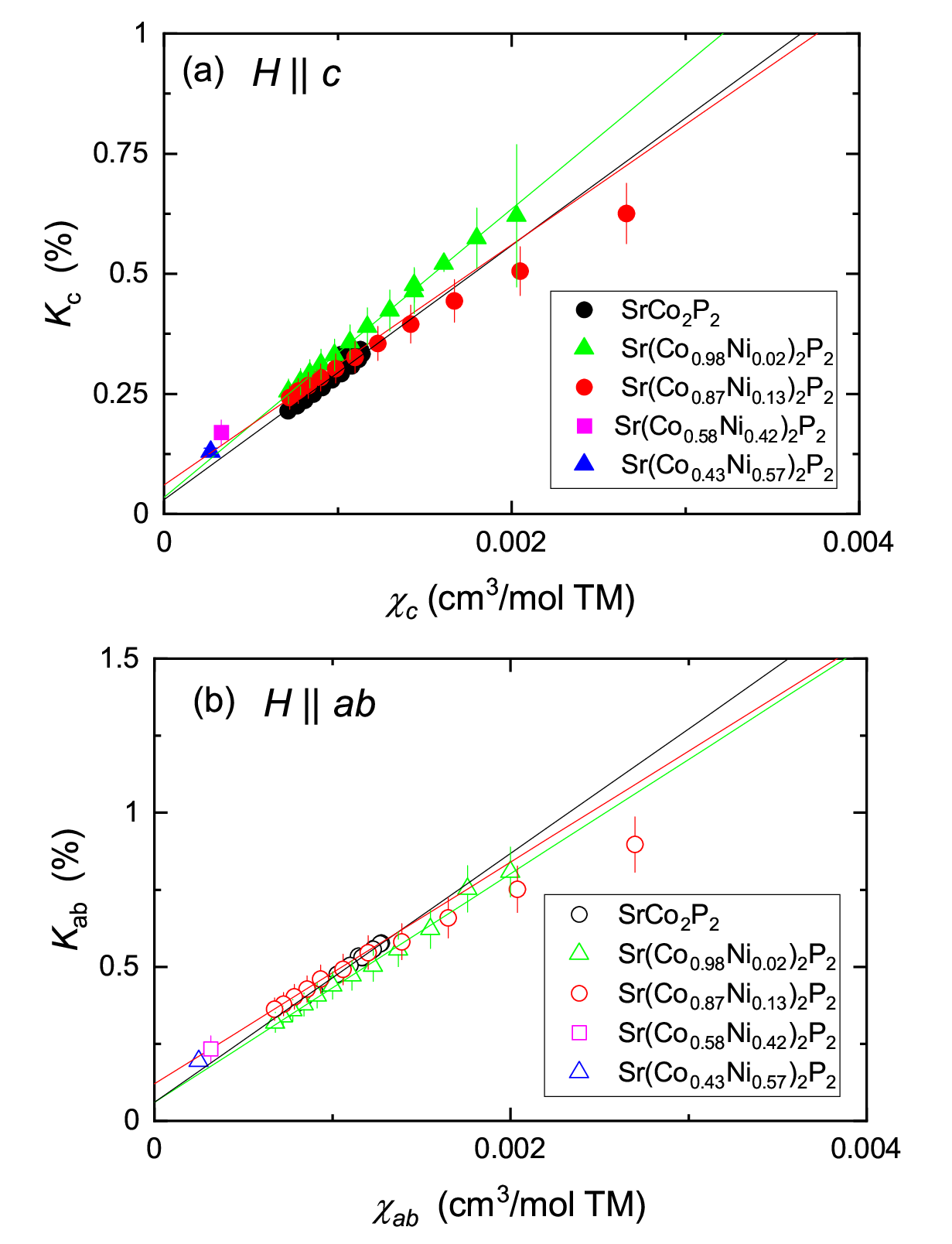}
\caption{NMR Knight shift $K$ versus magnetic susceptibility $\chi$ plots for the corresponding $c$ (a) and $ab$ (b) components of $K$  in Sr(Co$_{1-x}$Ni$_x$)$_2$P$_2$  with temperature as an implicit parameter. The solid lines are linear fits.  Note that as the NMR shifts for $x$ = 0.42 and 0.57 are nearly independent of temperature, we plot the representative points (the squares in majenta for $x$ = 0.42 and  the triangles in blue for $x$ = 0.57) in (a) and (b)  measured at  $T$ = 290-300 K showing that the points are on the lines and the values of $K_0$ are nearly independent of $x$. }
\label{fig:K-chi}
\end{figure}

    Figure \ref{fig:K} shows the temperature dependence of the NMR shift for $H\parallel c$~axis ($K_c$) and  $H\parallel ab$ plane ($K_{ab}$) for Sr(Co$_{1-x}$Ni$_x$)$_2$P$_2$. 
    For the case of $x$ = 0,  both $K_c$ and $K_{ab}$ show broad maxima around 25 K and 110 K, consistent with  the temperature dependence of the magnetic susceptibility shown in the top panel in the right column. 
These results are consistent with the previous NMR measurements reported by Imai ${\it et ~al.}$ \cite{Imai2017, Imai2014}. 
     For $x$ = 0.02 and 0.13,  both $K_c$ and $K_{ab}$ increase with decreasing temperature showing the Curie-Weiss-like behaviors, consistent with the temperature dependence of the magnetic susceptibility \cite{Schmidt2023}.     
    In contrast, a nearly temperature-independent behavior of  both $K_c$ and $K_{ab}$ is observed in $x$ = 0.42 by which substitution value  the AFM state has disappeared.  
   This indicates that the magnetic susceptibility of the dominant part of the crystal is nearly independent of temperature, although the magnetic susceptibility data show an increase with decreasing temperature. 
    Similarly, both $K_c$ and $K_{ab}$ for $x$ = 0.57 are nearly temperature independent expected for non-magnetic  metals,  although again the magnetic susceptibility data show small increases with decreasing temperature. 
  These results indicate that  the increase in $\chi$ might be due to the distribution of Ni content and/or  impurity spins . 


    The NMR shift $K$ is related to the local electron spin susceptibility  $\chi_{\rm spin}$ through the hyperfine coupling constant $A$ by  $K=K_0+\frac{A}{\mu_{\rm B}N_{\rm A}}\chi_{\rm spin}$, where $\mu_{\rm B}$ is the Bohr magneton, $N_{\rm A}$ is Avogadro's number, and $K_0$ is the temperature independent chemical shift originating from the core electrons of P atoms.  
    As shown in Fig. \ref{fig:K} (the right column),  the magnetic susceptibilities are nearly isotropic, although slight magnetic anisotropies can be seen below $\sim$ 200 K and 70 K for $x$ = 0 and 0.02, respectively. 
    Therefore the clear difference between $K_c$ and $K_{ab}$  is due to the different hyperfine coupling constants $A_c$ and $A_{ab}$ which can be estimated from the  so-called $K-\chi$ plot analysis.


     Figures \ref{fig:K-chi}(a)  and \ref{fig:K-chi}(b) plot $K_{c}$ and $K_{ab}$ against $\chi_{c}$ and $\chi_{ab}$, respectively, for all compounds with
$T$ as an implicit parameter where $T$ is chosen to be well above magnetic ordering temperatures for $x$ = 0.02 and 0.13 to avoid the effects of the magnetic ordering.
    $K_{ab}$ and $K_c$ vary with the corresponding $\chi$ as expected, although one can see a slight deviation from the linear relationship. 
    From the slopes of the fitted lines for the high-temperature regions above $\sim$ 100 K shown by the lines, we estimated the hyperfine coupling constants parallel to the $c$ axis $A_{c}= (1.47\pm 0.1) $ T/$\mu_{\rm B}$ and parallel to the $ab$ plane $A_{ab}= (2.25 \pm 0.1) $ T/$\mu_{\rm B}$ for $x$ = 0, $A_{c}= (1.67\pm 0.15) $ T/$\mu_{\rm B}$ and $A_{ab}= (2.07 \pm 0.15) $ T/$\mu_{\rm B}$ for $x$ = 0.02, and $A_{c}= (1.48\pm 0.12) $ T/$\mu_{\rm B}$ and $A_{ab}= (2.01 \pm 0.12) $ T/$\mu_{\rm B}$ for $x$ = 0.13. 
    Similar, but slightly different values of $A_{c}= 1.68 $ T/$\mu_{\rm B}$ and $A_{ab}= 2.66 $ T/$\mu_{\rm B}$  have been reported for $x$ = 0 \cite{Imai2017}.
    The values of chemical shift $K_0$  were also estimated to be $K_0$ = 0.06 (0.03) $\%$ for $H\parallel ab$ ($H\parallel c$) for $x$ = 0,  $K_0$ = 0.06 (0.035) $\%$ for $x$ = 0.02, and  $K_0$ = 0.12 (0.06) $\%$ for $x$ = 0.13.

\begin{figure}[t]
\centering
\includegraphics[width=\columnwidth]{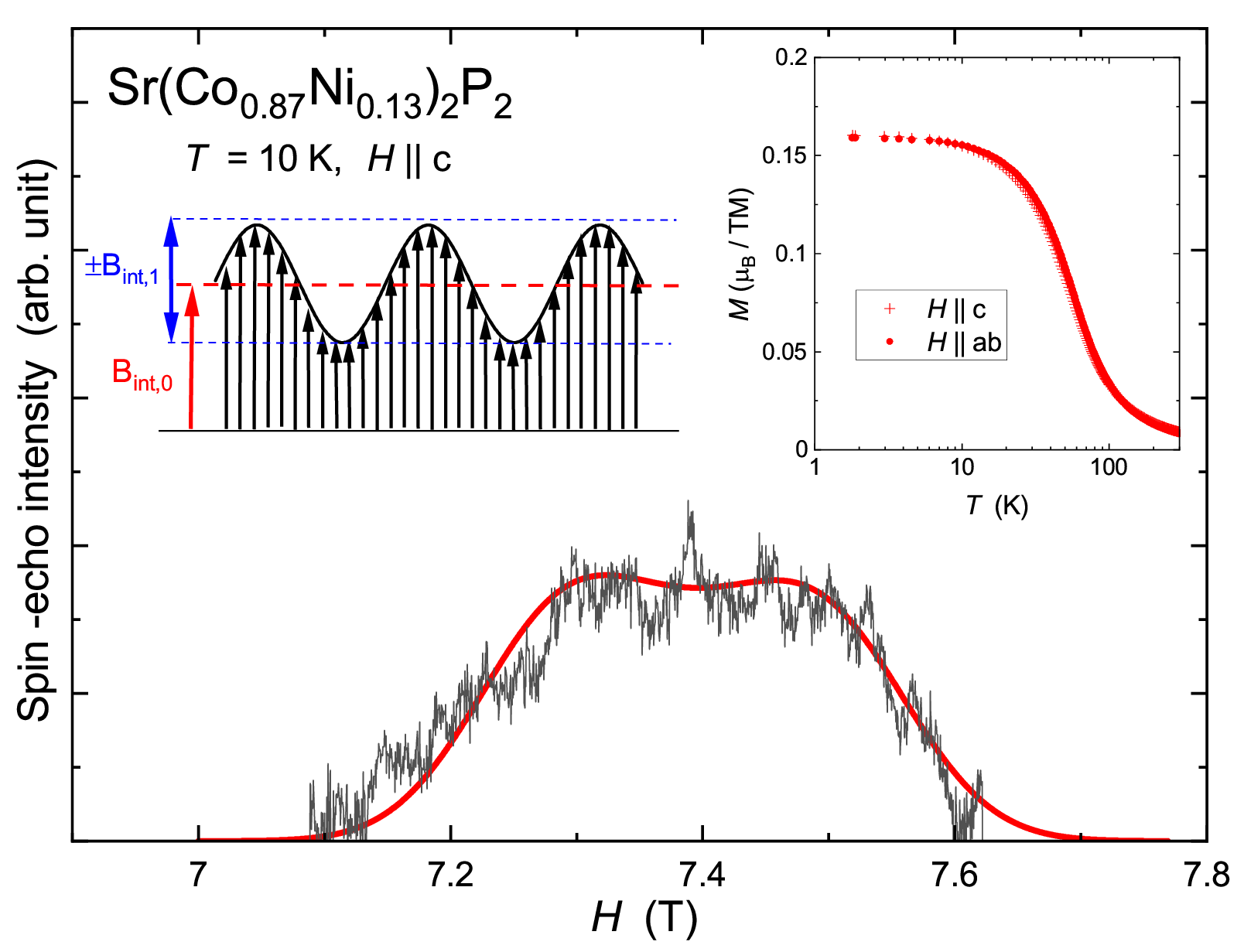}
\caption{Typical field-swept $^{31}$P-NMR spectrum  in the magnetically ordered state of $x$ = 0.13 measured at 10 K under $H\parallel c$. The red curve is the calculated spectrum based on the modulated magnetic ordered state illustrated in the left inset with  $B_{\rm int,0}$ = 0.22 T and $B_{\rm int,1}$ = 0.14 T (see the text for details). The arrows in the left inset represent the internal magnetic induction at the P sites.  The right inset shows the temperature dependence of magnetization of the crystal used in this study  for $H\parallel c$ and $H\parallel ab$ under $H$ = 7 T.   
}
\label{fig:Spectrum_OrderState}
\end{figure}

\subsection{$^{31}$P NMR spectrum in magnetically ordered state}

   As described above, we observed  characteristic rectangular-like shape spectra in the magnetically ordered states in $x$ = 0.02 and 0.13.
   For $x$ = 0.13, the antiferromagnetic ground state has been reported under a small magnetic field of 0.01 T \cite{Schmidt2023}.
   Although such a rectangular-like line shape is expected for powder samples of antiferromagnets \cite{Yamada1986, Devi2022}, this can not be applied here since we used single crystals.
   In addition, the AFM state in $x$  = 0.13 is transformed into a forced ferromagnetic (saturated paramagnetic) state under strong magnetic fields. 
   In fact, the magnetization measurements show a saturated magnetization of $\sim$ 0.15 $\mu_{\rm B}$ per TM ions above 1.5 T under $H\parallel ab$ at 2 K and a similar saturated magnetization of  $\sim$ 0.15  $\mu_{\rm B}$ is observed above 3 T for $H\parallel c$ \cite{Schmidt2023}. 
   This indicates that all magnetic moments in the magnetic ordered state are aligned along the magnetic field direction.
   This can also be seen in the temperature dependence of magnetization $M$ of the $x$ = 0.13 crystal used in the present study at $H$ = 7 T shown 
in Fig. \ref{fig:K} (right middle panel) and replotted in the right inset of Fig. \ref{fig:Spectrum_OrderState} on a semi-log scale  where $M$ shows a saturation moment of 0.16 $\mu_{\rm B}$/TM below $\sim$ 10 K with no obvious magnetic anisotropy.

    Therefore, as our NMR spectrum measurements have been performed above $H$ = 7 T,  we consider that the NMR spectra were measured in the state where the magnetic moments are saturated in the  $x$ = 0.13 compound by the application of the magnetic field. 
 The broad NMR line is mainly due to the distribution of the hyperfine field at the P site due to the Ni substitution which generates different environments of the P site with respect to  the number of nearest neighbors (NN) of a given Co occupied by Ni. From the estimate of the probability of finding $N$ (= 1-4) Ni at NN  Co sites with $P(N,x)$ =   $C_N^4x^N(1-x)^{4-N}$,  we found that only the distribution of the hyperfine field cannot explain the rectangular-like shape of the spectrum. 
 Therefore, one needs to introduce a distribution of Co ordered moments in order to explain the characteristic shape of the NMR line.
    Thus we conclude that the magnitude of the Co-ordered moments is not uniform and the magnetic states are inhomogeneous in the magnetic ordered state of the Ni-substituted SrCo$_2$P$_2$.
    To explain the observed spectra, we modeled that the Co ordered moments are not uniform and the magnitude is modulated with a sinusoidal function as shown in the inset of Fig. \ref{fig:Spectrum_OrderState} which produces the uniform and distributed  internal magnetic inductions    $B_{\rm int,0}$  and $B_{\rm int,1}$ at the P site, respectively.  
    Here it is noted that the modulation of the magnetic moment is assumed to be incommensurate with the lattice.   
 
   The red curve in Fig. \ref{fig:Spectrum_OrderState} is a calculated result with $B_{\rm int,0}$ = 0.22 T and $B_{\rm int,1}$ = 0.14 T with an appropriate inhomogeneous magnetic  broadening,  which reasonably reproduces  the observed spectrum (shown by the black curve) measured at $T$ = 10 K  with $H\parallel c$ for Sr(Co$_{0.87}$Ni$_{0.13}$)$_2$P$_2$ ($x$ = 0.13). 
   Thus we consider that the rectangular-like NMR spectra observed in the magnetically ordered state are due to the variation of the Co-ordered moments.        
     Utilizing the values of $A_c$ = 1.48~T/$\mu_{\rm B}$, the uniform and the distribution components  of the magnetic moment are estimated to be 0.15$\pm$0.01  and 0.10$\pm$0.01 $\mu_{\rm B}$, respectively.  
The value of the uniform magnetic moment is in great agreement with  a value of 0.16   $\mu_{\rm B}$ estimated from the $M$ measurements shown in the right inset of Fig. \ref{fig:Spectrum_OrderState}. 

     The red curves in the left column of Fig. \ref{fig:T-dep_spectrum}(c) show calculated results where  $B_{\rm int,0}$($B_{\rm int,1}$)  = 0.22(0.14)  T are nearly independent of temperature below 10 K, and decrease gradually above 10 K to 0.19(0.11) T at 20 K and 0.17(0.10) T at 30 K for $H\parallel c$.
    The temperature-dependent behavior is consistent with the temperature dependence of $M$ shown in 
the inset of Fig. \ref{fig:Spectrum_OrderState}.

     For the case of $H\parallel ab$ [the right column of Fig.~\ref{fig:T-dep_spectrum}(c)], the calculated spectra also capture the characteristic shape of the observed spectra where we used $B_{\rm int,0}$($B_{\rm int,1}$)  = 0.33(0.19), 0.31(0.20), 0.25(0.16), and 0.20(0.13) T for $T$ = 4.2, 10, 20, and 30 K, respectively, indicative of the variation of the  Co ordered moments in the magnetically ordered state. 
    The uniform and the distribution components of the Co ordered moments for the $H\parallel ab$ direction are estimated using $A_{ab}$ = 2.01 T/$\mu_{\rm B}$ to be 0.16$\pm$0.01 and 0.10$\pm$0.01 $\mu_{\rm B}$ at 4.2 K.
   These values are very close to those for $H\parallel c$. 
    Again the estimated uniform magnetization from the NMR spectrum simulation is in excellent agreement with the magnetization data, indicating that the magnetization  measurements picked up the average value of magnetization as expected.  

     It is noted that the model can also roughly reproduce the NMR spectra at low temperatures for $x$ = 0.02 as shown by the red curves in Fig. \ref{fig:T-dep_spectrum}(b), although the tails of the spectra at the lower magnetic field side cannot be well reproduced. 
  This is presumably due to the large inhomogeneous distribution of hyperfine field coming from the Ni substitution as described above.
  For example, $B_{\rm int,0}$($B_{\rm int,1}$) = 0.082(0.037) T for $H\parallel c$ and  0.12(0.065) T for $H\parallel ab$ were used for the calculation for the spectra at 1.6 K. 
   By using $A_c(A_{ab})$ = 1.67(2.07) T/$\mu_{\rm B}$, the uniform and distribution components of the Co ordered moments are estimated to be  0.049(0.022)  and 0.058(0.031) $\mu_{\rm B}$ for $H\parallel c$ and $H\parallel ab$, respectively, for $x$ = 0.02.
   These are also in good agreement with 0.049 and 0.056 $\mu_{\rm B}$ for for $H\parallel c$ and $H\parallel ab$ determined by the magnetization measurements under 7 T shown in the right panel in Fig. \ref{fig:K} of the measured crystal of $x$ = 0.02.
  Thus we conclude that the magnetic states in Sr(Co$_{1-x}$Ni$_x$)$_2$P$_2$ are non-uniform  and the distribution of the Co-ordered moments is incommensurate with the lattice.  
  
We have tried to measure the $^{31}$P NMR spectrum around  $H$ = 1 T in the AFM state for $x$ = 0.13. 
However, the signal intensity was too poor to measure, so we were not able to obtain the spectrum in the AFM state.   
Therefore, we cannot discuss the magnetic structure in the AFM state in Sr(Co$_{1-x}$Ni$_x$)$_2$P$_2$.
Nevertheless, it is worth to mention that, in the isostructural As-system Sr(Co$_{1-x}$Ni$_x$)$_2$As$_2$,  the AFM state appearing in $x$ = 0.013-0.25 with a maximum $T_{\rm N}$ = 26.5 K at $x$ = 0.10 has been determined to be a planar helical magnetic state originating from FM interactions in the $ab$ plane and weaker AFM interactions along the helix $c$ axis \cite{Sangeetha2019,Wilde2019}. 
   As we describe below, this magnetic structure would be consistent with our observation of strong ferromagnetic spin fluctuations in $x$ = 0.13. 
Another possible magnetic structure is the A-type antiferromagnetic state observed in CaCo$_2$P$_2$  \cite{Reehuis1998,Imai2017,Baumbach2014PRB}. 
  Further experiments such as ND measurements are highly required to determine the magnetic structure in the AFM state in  Sr(Co$_{1-x}$Ni$_x$)$_2$P$_2$.

\subsection{$^{31}$P spin-lattice relaxation time $T_1$}

\begin{figure}[t]
\centering
\includegraphics[width=\columnwidth]{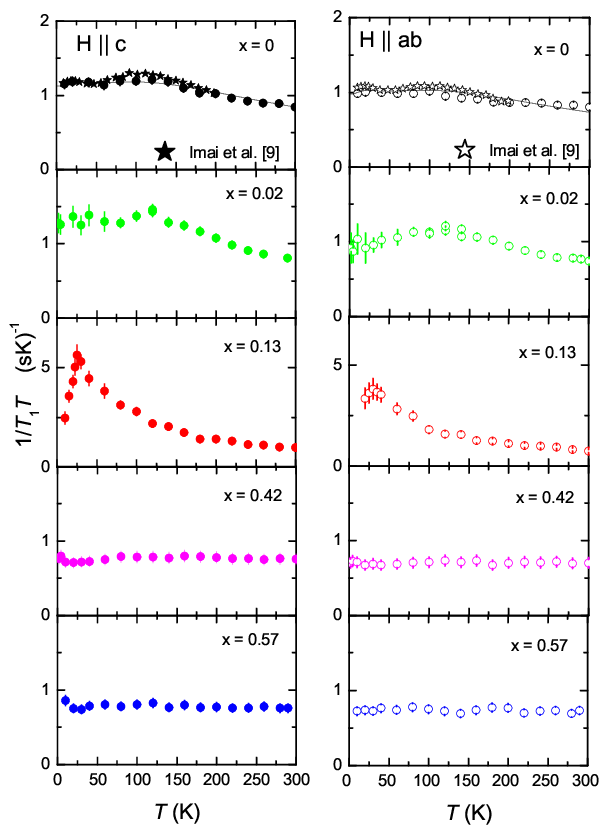}
\caption{ Temperature and $x$ dependences of the $^{31}$P spin-lattice relaxation rate divided by temperature (1/$T_1T$) for both field directions  [$H\parallel c$ (left panels),  $H\parallel ab$ (right panels)]   in Sr(Co$_{1-x}$Ni$_x$)$_2$P$_2$. The 1/$T_1T$ data for $x$ = 0 shown by closed and open stars are from Ref. \cite{Imai2017}.  The black solid curves for $x$ = 0 are calculated results with the model discussed in the text. Note that the vertical scales for $x$ = 0.13 are different from others.}
\label{fig:T1}
\end{figure}

\begin{figure}[t]
\centering
\includegraphics[width=\columnwidth]{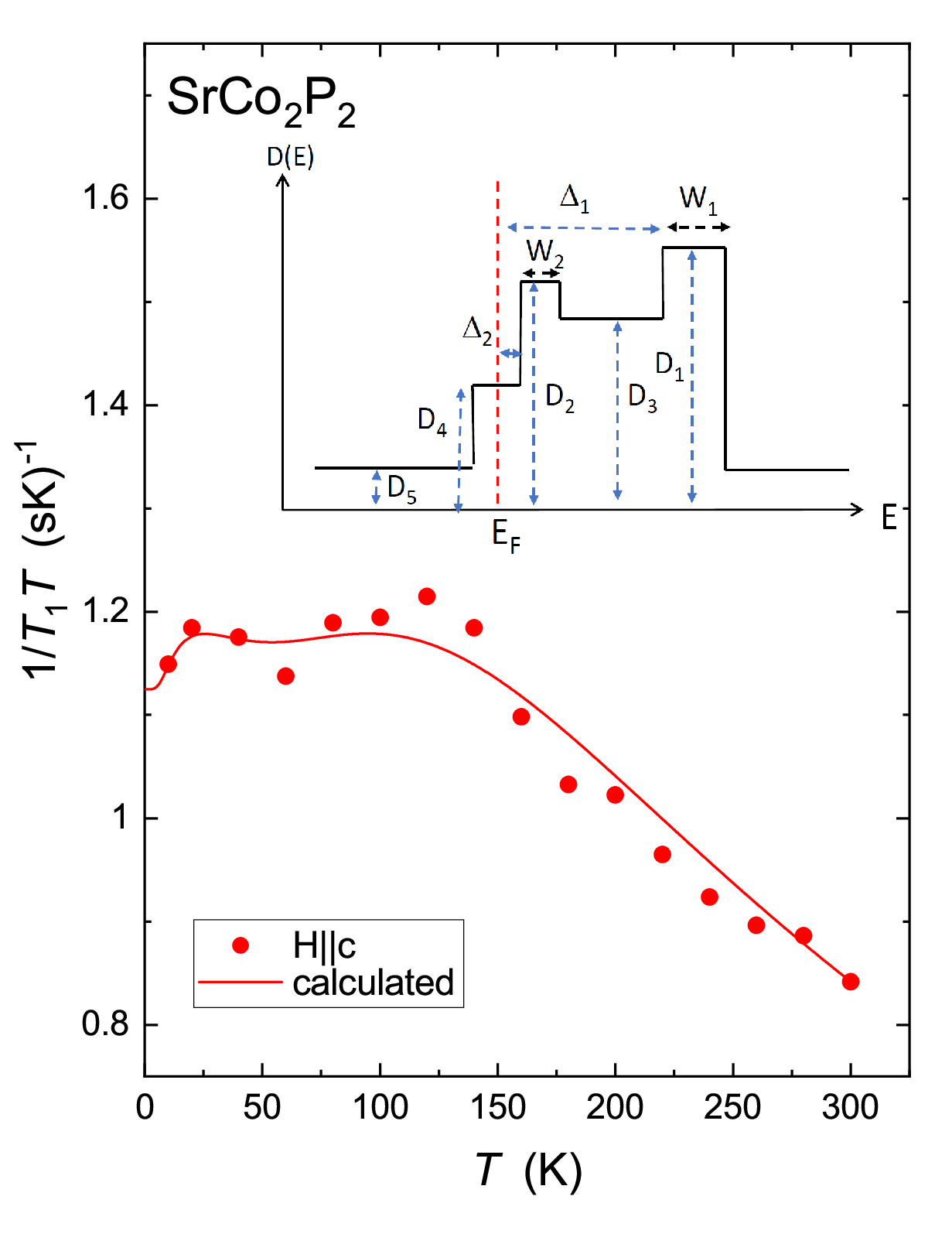}
\caption{$T$ dependence of the $^{31}$P spin-lattice relaxation rate divided by temperature (1/$T_1T$) for $H\parallel c$ of SrCo$_2$P$_2$. The red curve is the calculated result based on the the double peak structure in the density of states  near Fermi energy ($E_{\rm F}$)  shown in the inset.}
\label{fig:T1_analysis}
\end{figure}

  To investigate the dynamical magnetic properties, we have measured 1/$T_1$ versus $T$. 
Figure \ref{fig:T1} shows the temperature and $x$ dependence of  1/$T_1T$ for both field directions (left: $H\parallel c$, right: $H\parallel ab$)  in Sr(Co$_{1-x}$Ni$_x$)$_2$P$_2$. 
   At $x$ = 0, similar to the case of $K$, and thus $\chi$,  broad maxima are observed around $\sim$ 25 K and 120 K, as reported previously \cite{Imai2017, Imai2014}.
 For comparison, we show the 1/$T_1T$ data for $x$ = 0  reported from Imai {\it et~al.} \cite{Imai2017}  by closed and open stars in the figures, which show a relatively good agreement with our results. 
   With Ni substitution of $x$ = 0.02,  1/$T_1T$ still shows a similar temperature dependence with those of 1/$T_1T$ for $x$ = 0 exhibiting a broad maximum around 120 K for both magnetic field directions where the broad maximum around 25 K is not clearly observed.  
 No apparent anomaly due to ferromagnetic phase transition is observed around $T_{\rm C}$, which is probably broadened into a high temperature cross over into the saturated paramagnetic state by the application of the magnetic field of $\sim$ 7.4 T. 
   On the other hand, at $x$ = 0.13, 1/$T_1T$ is enhanced and shows clear peaks around $T_{\rm N}$ for both magnetic field directions. 
   However, the peaks in 1/$T_1T$  should  not be due to the  AFM transition because the ground state of the compound is a forced ferromagnetic (saturated paramagnetic) state under $H$ =  $\sim$ 7.4 T. It will be interpreted as a crossover from the paramagnetic to the forced ferromagnetic states. 
   With further Ni substitution of $x$ = 0.42 and to 0.57, in contrast, the values of 1/$T_1T$ are suppressed and 1/$T_1T$ follows the Korringa law, $T_1T$ = const. which is indicative of a paramagnetic metallic state without significant magnetic fluctuations.   

   First, we analyze the characteristic temperature dependence of 1/$T_1T$ showing the double-maxima observed in $x$ = 0.  
Since the characteristic temperature dependence of 1/$T_1T$ can be due to a band structure near the Fermi energy $E_{\rm F}$,   we calculated it based on a  model where we adopt a double-peak structure of the density of states (DOS) shown in the inset of Fig. \ \ref{fig:T1_analysis}. 
    
      In this model, the Fermi energy ($E_{\rm F}$) is assumed to be just below the double peaks having the widths of $W_1$ and $W_2$ away from the $E_{\rm F}$  by $\Delta_1$ and $\Delta_2$, respectively. 
      ${\cal D}_i$ ($i=1, 2 , 3, 4, 5)$ represent the DOS for the band structure near $E_{\rm F}$. 
      Using the formula, 
\begin{equation}
\frac{1}{T_1}\sim\int_0^\infty{\cal D}^2(E)f(E)(1-f(E)){\rm{d}}E 
\label{eq:T1}
\end{equation}
where $f(E)$ is the Fermi distribution function, we calculated $1/T_1T$ with a set of parameters of $\Delta_1$ = 300 K, $\Delta_2$ = 25 K, $W_1$ = 125 K, $W_2$ = 30 K, $D_i/D_1$  = 1, 0.82, 0.75, 0.55,  and 0.15 for $i$ = 1, 2, 3, 4, and 5, respectively.
The calculated result reasonably reproduces the experimental data, as shown by the red curve in Fig. \ \ref{fig:T1_analysis}.  
   This indicates that the characteristic temperature dependence of $1/T_1T$ exhibiting double maxima can be explained by the rigid  peculiar band structure modeled. 
   The calculated results are also shown in Fig. \ref{fig:T1}. 
It should be noted that, since the number of  parameters is large (9 parameters)  in the model, we cannot precisely determine the value of each parameter from the fitting. Therefore, there is relatively large uncertainty in the value of each parameter ($\sim$ 25 \%); however, again we point out  that the model of the double-peak structure of the DOS qualitatively explains the characteristic temperature dependence of 1/$T_1T$. 
   It is interesting to point out  that such a double-peak DOS structure in SrCo$_2$P$_2$ has been reported from the density-function-theory band-structure calculations  which reproduces the experimental data of de Haas-van Alphen effect experiments \cite{Gotze2021}.
 According to the calculations, the positions of the double peaks in the DOS are largely influenced by the crystal structure parameters such as lattice parameters and the Wyckoff position of P, and also different approximations used in the calculations \cite{Gotze2021}. Nevertheless, it is interesting to point out that the centers of the calculated two peaks are located around 0.005-0.011 eV and 0.02-0.035 eV above $E_{\rm F}$  which are not far from the values obtained by the fit with  $\Delta_2$ = 25 K (0.0022 eV) and $\Delta_1$ = 300 K (0.026 eV). In contrast, it is noted that the widths of each peak from the theoretical calculations are roughly estimated to be $\sim$ 0.011 eV and $\sim$ 0.005 eV for the peak around 0.005-0.011 eV and 0.02-0.035 eV, respectively, which are  different from our fitting values of $W_2$ = 30 K (0.0026 eV) and $W_1$ = 125 K (0.011 eV).

   We also calculated the temperature dependence of Knight shift using  the same band structure model.   
    The solid curves in Fig.  \ \ref{fig:K} (the top panel on the left) are the calculated results with  appropriate proportional constants utilizing  the same set of parameters used for the $1/T_1T$ fitting except for the small change in  ${\cal D}_4$/${\cal D}_1$  from 0.55 to 0.45. 
    Although the fits do not reproduce the data perfectly, the model seems to capture the behavior of $K$ qualitatively.
    The reason for the slight change in ${\cal D}_4$ is not clear,  but it may be due to magnetic fluctuations which do not follow the temperature dependence of the Knight shift.

   From the calculations, it turns out that, as naturally expected, the broad maximum around 125 K in $1/T_1T$  is mainly due to the peak in the DOS far away from the $E_{\rm F}$ and the maximum around 25 K originates from the peak close to the $E_{\rm F}$. 
   Therefore, the similar  temperature dependence of 1/$T_1T$ observed in $x$ = 0.02 suggests that the DOS structure is not modified much by the Ni substitution although the system shows the ferromagnetic ground state. 
 At present we do not have a clear idea to explain this. However, it may suggest that a very small change in DOS leads to the ferromagnetic long-range magnetic ordering and SrCo$_2$P$_2$ is located close to ferromagnetic instability. 
    In contrast, for $x$ = 0.13, 1/$T_1T$ are enhanced, which cannot be explained  by the simple band model. 
    For $x$ = 0.42 and 0.57, the 1/$T_1T$ = const. behavior can be explained by a flat structure of the DOS near the $E_{\rm F}$ expected for a normal metallic state.  
    This indicates that the Ni substitution greater than $x$ = 0.42 completely smears out the double-peak structure, leading to a more itinerant nature of Co 3d electrons.

\subsection{Magnetic fluctuations}

\begin{figure}[t]
\centering
\includegraphics[width=\columnwidth]{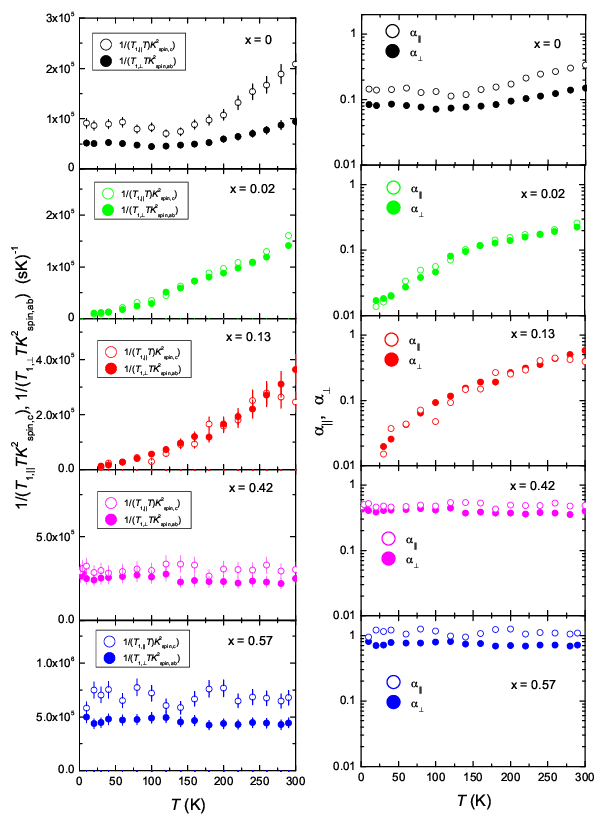}
\caption{Left panels: Temperature dependence of the Korringa ratios $1/T_{1,\bot}TK_{\text{spin},ab}^2$ (closed symbols) and
$1/T_{1,\|}TK_{\text{spin},c}^2$ (open symbols) for spin fluctuations in the $ab$ plane and along the $c$ axis, respectively, in Sr(Co$_{1-x}$Ni$_x$)$_2$P$_2$.
  Right panels: Semi-log plot of the temperature  dependence of the parameter $\alpha_{\bot}$ for spin fluctuations in the $ab$ plane (closed symbols) and $\alpha_{\|}$ along the $c$ axis (open symbols). }
\label{fig:T1TK2}
\end{figure}

      Now we discuss the magnetic fluctuations in Sr(Co$_{1-x}$Ni$_x$)$_2$P$_2$ based on  a modified Korringa relation analysis using the $T_1$ and the electron spin part of Knight shift ($K_{\rm spin}$) data.
    Within a Fermi liquid picture, $1/T_1T$ is proportional to the square of the DOS near the Fermi energy ${\cal D}(E_{\rm F})$ and $K_{\text{spin}} (\propto \chi_{\text{spin}}$) is proportional to ${\cal D}(E_{\rm F})$. 
    In particular, $T_1TK_{\text{spin}}^2$  = $\frac{\hbar}{4\pi k_{\rm B}} \left(\frac{\gamma_{\rm e}}{\gamma_{\rm N}}\right)^2$ = $S$, which is the Korringa relation.  
    Deviations from $S$ can reveal information about electron correlations in materials \cite{Moriya1963,Narath1968}, which are expressed via the parameter $\alpha=S/(T_1TK_{\text{spin}}^2)$. 
      For instance, the enhancement of $\chi(\mathbf{q}\neq 0)$ increases $1/T_1T$ but has little or no effect on $K_{\text{spin}}$, which probes only the uniform $\chi$ with $\mathbf{q}$ = 0.  
    Thus  $\alpha >1$ for AFM correlations and $\alpha <1$ for FM correlations.

     Since 1/$T_1T$ probes magnetic fluctuations perpendicular to the magnetic field,  it is natural to consider the Korringa ratio 1/($T_{1, \perp}TK^2_{{\rm spin}, ab}$), where 1/($T_{1,\perp}T$) = 1/$(T_1T)_{H||c}$, when examining the character of magnetic fluctuations in the $ab$ plane   \cite{Wiecki2015-2}. 
     We also consider the Korringa ratio 1/($T_{1, ||}TK^2_{{\rm spin}, c}$) for magnetic fluctuations along the $c$ axis. 
     Here, 1/($T_{1, \parallel}T$) is estimated from 2/$(T_{1}T)_{H||ab}$~$-$~$1/(T_{1}T)_{H||c}$   \cite{Wiecki2015-2}.

     The temperature and $x$ dependences of 1/($T_{1, \perp}TK^2_{{\rm spin}, ab}$)  and  1/($T_{1, ||}TK^2_{{\rm spin}, c}$)  above the magnetic ordering temperatures in Sr(Co$_{1-x}$Ni$_x$)$_2$P$_2$ are shown in the left panels of Fig.~\ref{fig:T1TK2}.  
     The calculated $\alpha_\parallel$ and $\alpha_\perp$ are shown in the right panels of Fig.~\ref{fig:T1TK2}.
  
   For $x$ = 0, $\alpha_\parallel$ ($\alpha_\perp$)  decrease from $\sim$0.4 (0.2) at 300 K to $\sim$0.08 (0.14) around 110 K and then become nearly temperature independent below that temperature, indicating dominant ferromagnetic spin correlations between Co spins at low temperatures  in the compound.
    Furthermore, $\alpha_\perp$ is less than $\alpha_\parallel$ suggesting the in-plane ferromagnetic fluctuations are stronger than those along the $c$ axis. 
    The lowest values of  $\alpha_\parallel$ and $\alpha_\perp$  are almost comparable, or slightly greater than those isostructural compounds BaCo$_2$As$_2$ and SrCo$_2$As$_2$ in which dominant ferromagnetic spin fluctuations have also been reported \cite{ Wiecki2015,BaCo2As2}.
   It is interesting to point out that the coexistence of ferromagnetic and antiferromagnetic spin fluctuations have been reported in SrCo$_2$As$_2$ \cite{Wiecki2015, Li2019}, although we do not have any clear evidence from the present $^{31}$P-NMR measurements  for  AFM spin fluctuations in SrCo$_2$P$_2$.

For $x$ = 0.02, although the values of $\alpha_\parallel$ and $\alpha_\perp$ are nearly the same with the $\alpha_\perp$ value (but slightly smaller than $\alpha_\parallel$)   in $x$ = 0 at 300 K , $\alpha_\parallel$ and $\alpha_\perp$ decrease rapidly down to 0.015, almost one order in magnitude less than the case of $x$ = 0, indicating a strong enhancement of ferromagnetic spin fluctuations at low temperatures in $x$ = 0.02. 
It is also interesting to note that the ferromagnetic spin fluctuations are nearly isotropic in $x$ = 0.02.  

A similar enhancement of the isotropic ferromagnetic spin fluctuations at low temperatures is also observed in $x$ = 0.13.  
The values of $\alpha_\parallel$ and $\alpha_\perp$ for $x$ = 0.13 are almost comparable with those in $x$ = 0.02. 
These results mean that the strong ferromagnetic spin fluctuations are induced with only a 2 \% Ni substitution and are kept almost the same up to $x$ = 0.13 where the magnetic ground state changes from ferromagnetic ($x$ = 0.02) to antiferromagnetic ($x$ = 0.13).
It is also interesting to point out that we do not observe any trace of AFM spin fluctuations in $x$ = 0.13, indicating that the magnetic fluctuations in $x$ = 0.13 are dominated by the ferromagnetic fluctuations.
   At present, the reason for no observation of  the AFM spin fluctuations  in $x$ = 0.13 is not well understood. 
   However,  it is interesting to point out that, in the A-type antiferromagnet CaCo$_2$P$_2$ where the Co moments are ferromagnetically aligned in the $ab$ plane and the moments adjacent along the $c$ axis are aligned antiferromagnetically \cite{Reehuis1998,Imai2017,Baumbach2014PRB}, no clear AFM spin fluctuations were observed and the magnetic fluctuations were found to be dominated by ferromagnetic ones \cite{Higa2018}.
   In addition, as described above, another possible AFM state of the planar helical magnetic state observed in Sr(Co$_{1-x}$Ni$_x$)$_2$P$_2$ \cite{Sangeetha2019, Wilde2019}  would also provide dominant ferromagnetic spin fluctuations. 
    Although the AFM structure in Sr(Co$_{1-x}$Ni$_x$)$_2$P$_2$ is still an open question, such  magnetic states  might lead to the dominant ferromagnetic spin fluctuations.


With Ni substitution of $x$ = 0.42, $\alpha_\parallel$ and $\alpha_\perp$  increase and become temperature independent without showing any indication of quantum critical behavior. 
At $x$ = 0.57,  the nearly temperature independent $\alpha_\parallel$ and $\alpha_\perp$ become  closer to unity, indicating that the magnetic fluctuations disappeared in the compound.
   Here we tentatively used $K_0$ = 0.03 (0.06) for $H\parallel c$ ($H\parallel ab$) since the data points for $x$ = 0.42 and 0.57 in the $K=\chi$ plots [Figs. \ref{fig:K-chi}(a) and (b)] are well on the fitting lines showing the nearly $x$ independence of $K_0$.


\begin{figure}[t]
\centering
\includegraphics[width=\columnwidth]{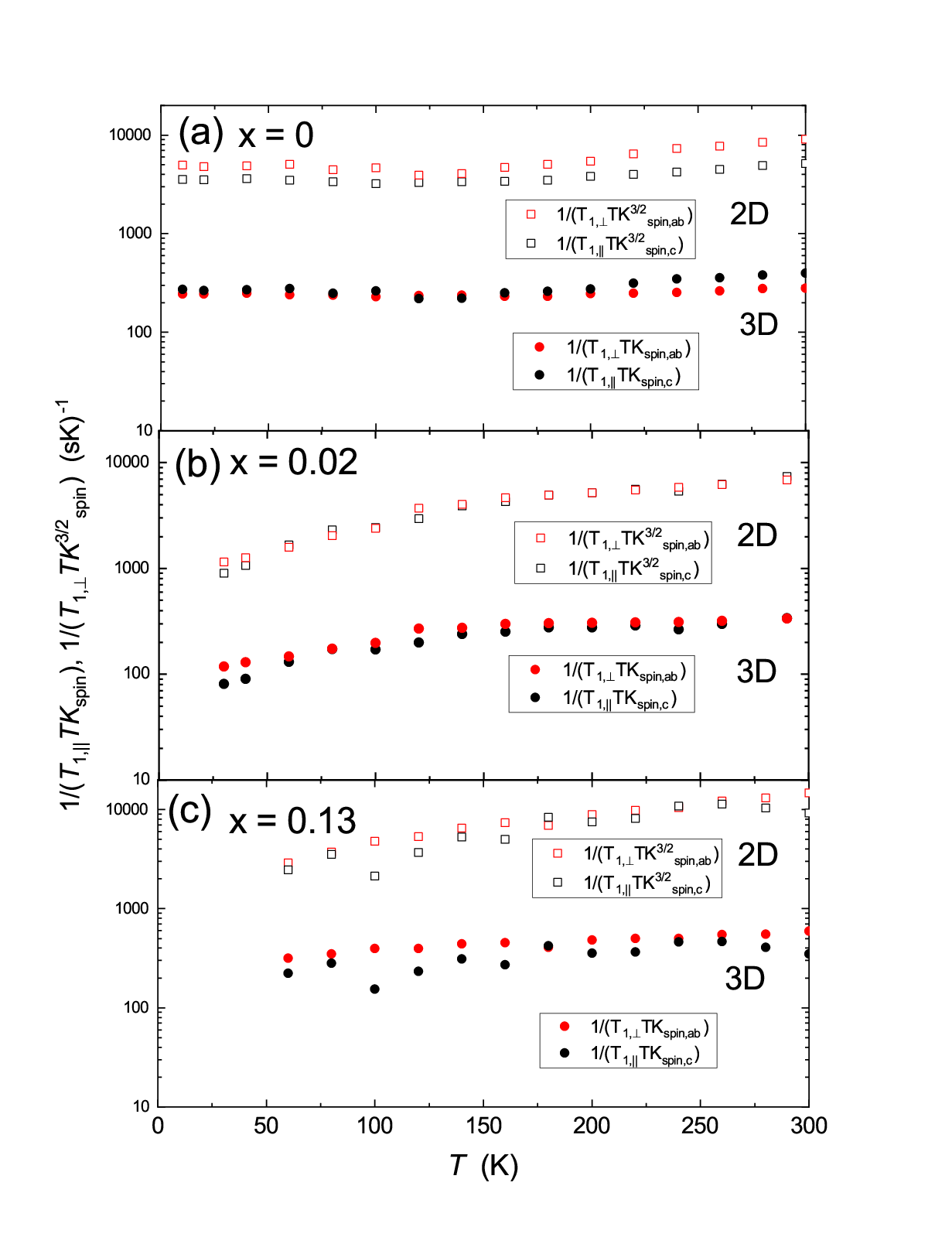}
\caption{ Temperature dependences of $1/T_{1}TK_{\text{spin}}$ (closed symbols) and $1/T_{1}TK_{\text{spin}}^{3/2} $ (open symbols) for the  in-plane (red) and the $c$ axis (black) directions for $x$ = 0 (a), 0.02 (b), and 0.13 (c). }
\label{fig:SCR}
\end{figure}

   It should be noted that, however, the Korringa analysis usually applies to paramagnetic materials where electron-electron interaction is weak.  
    Since  the compounds with $x$ = 0.02 and 0.13  exhibit magnetic orders in contrast to $x$ = 0, 0.42 and 0.57,  we also analyze NMR data based on self-consistent renormalization (SCR) theory, as Imai ${\it et~ al.}$ have performed in (Sr$_{1-x}$Ca$_x$)Co$_2$P$_2$ \cite{Imai2017}.
    As shown above, the magnetic fluctuations are governed by ferromagnetic spin correlations. 
In this case, according to SCR theory for weak itinerant ferromagnets,  1/$T_1T$ is proportional to $K_{\rm spin}$ or to $K^{3/2}_{\rm spin}$ for three dimensional (3D) or  two-dimensional (2D) ferromagnetic spin fluctuations, respectively \cite{SCR1, SCR2}. 
 
    Figures~\ref{fig:SCR}(b) and \ref{fig:SCR}(c) show  the $T$ dependence of 1/($T_1TK_{\rm spin})$ and 1/($T_1TK_{\rm spin}^{3/2})$ for the two directions for $x$ = 0.02 and 0.13, respectively, together with those in $x$ = 0 (a).
  For both $x$ = 0.02 and 0.13, it seems that both the 1/($T_{1,\parallel}TK_{\rm spin,c}$) and 1/($T_{1,\perp}TK_{{\rm spin},ab}$)  are nearly constant in the temperature region of 150 - 300 K while both  the 1/($T_{1,\parallel}TK_{{\rm s},c}^{3/2}$) and 1/($T_{1,\perp}TK_{{\rm s},ab}^{3/2}$) slightly increase with increasing temperature in that temperature region.
  This suggests  the ferromagnetic spin fluctuations are most likely characterized by 3D  in nature in the high-temperature region for both $x$ = 0.02 and 0.13. 
   It is noted that both 1/($T_1TK_{\rm spin})$ and 1/($T_1TK_{\rm spin}^{3/2})$ decrease with decreasing temperature below $\sim$150 K, suggesting that the SCR theory may not be applicable  to describe the dynamical properties of  the Co moments in $x$ = 0.02 and 0.13 exhibiting long-range magnetic ordered states. 
As for $x$ = 0, although the SCR theory for the 3D FM model seems slightly better than that for the 2D model, it is difficult to conclude the dimensionality of the FM fluctuations from the present NMR data.  

\begin{figure}[t]
\centering
\includegraphics[width=\columnwidth]{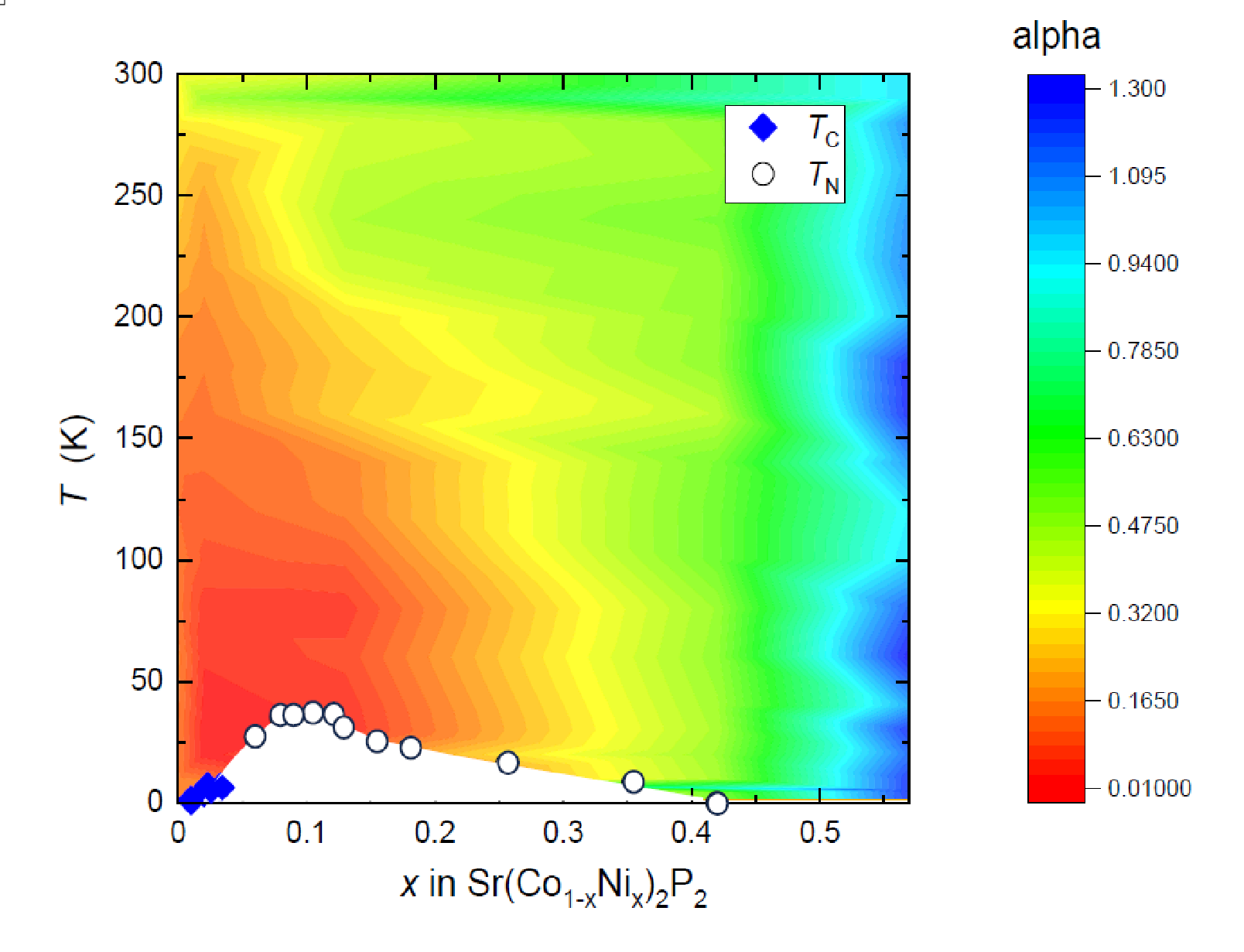}
\caption{ Contour plot of  $\alpha_\parallel$ vs temperature showing the magnitude of the spin fluctuations together with the magnetic phase transition temperatures from Ref. \cite{Schmidt2023}. Note that the lower the magnitude of $\alpha_\parallel$,  the stronger in the ferromagnetic spin fluctuations.   The blue diamonds and white circles represent $T_{\rm C}$ and $T_{\rm N}$, respectively, from Ref. \cite{Schmidt2023}. Note that, whereas the phase transition temperature data \cite{Schmidt2023}  were inferred from low field (zero field) data, the $\alpha_\parallel$ data were inferred from $\sim$ 7.4 T data.
The values of $\alpha_\parallel$ are not shown  in the magnetically ordered state (the white area).    }
\label{fig:Phase_Diagram}
\end{figure}

\section{Summary}
  In conclusion, we conducted $^{31}$P NMR measurements on Sr(Co$_{1-x}$Ni$_x$)$_2$P$_2$  to investigate its static and dynamic magnetic properties. 
    The characteristic  temperature dependences of  $K$  and 1/$T_1T$ exhibiting the double-maxima around 25 K and 125 K in the Stoner-enhanced paramagnetic metal  SrCo$_2$P$_2$  were reasonably explained by a simple model where a double-peak structure of the density of states near the Fermi energy is assumed.  
   We analyzed our $T_1$ and $K$ data based on the modified Korringa relation and discussed the evolution of magnetic fluctuations in the paramagnetic state  in Sr(Co$_{1-x}$Ni$_x$)$_2$P$_2$.  
   Figure \ref{fig:Phase_Diagram} summarizes   the evolution of the magnetic fluctuations where  the $x$ dependence of the Korringa ratios of the longitudinal magnetic fluctuation $\alpha_\parallel$ vs. temperature is shown in the contour plot together with the magnetic phase transition temperatures from Ref. \cite{Schmidt2023}. 
     Here we tentatively plotted the data of  $\alpha_\parallel$, but the in-plane magnetic fluctuations  $\alpha_\perp$ provide qualitatively the same behavior \cite{Korringa2}.  
 The lower values in the magnitude of $\alpha_\parallel$, much less than unity,  qualitatively indicate the stronger ferromagnetic spin fluctuations.   
  For $x$ = 0, $\alpha_\parallel$  is found to be much less than unity (the red-ish region in  Fig. \ref{fig:Phase_Diagram}) evidencing that  electron correlations are enhanced  around FM wavenumber $q$ = 0  in the paramagnetic  SrCo$_2$P$_2$. 
   Such FM spin fluctuations are enhanced with only 2 \% Ni substitution in the ferromagnetic Sr(Co$_{0.98}$Ni$_{0.02}$)$_2$P$_2$. 
   Similar strong FM spin fluctuations are observed in the antiferromagnet Sr(Co$_{0.87}$Ni$_{0.13}$)$_2$P$_2$ with $T_{\rm N}$ = 30 K.  
   With further Ni substitution up to 0.42 and to 0.57, $\alpha_\parallel$ gradually increases and becomes  close to unity (blue region in Fig. \ref{fig:Phase_Diagram})  indicating that such FM spin fluctuations completely disappeared, leading to a non-correlated metal.
   It is noted that no clear quantum critical behavior in Sr(Co$_{1-x}$Ni$_x$)$_2$P$_2$  can be observed around $x$ = 0.42 where the long-range magnetic state  is not observed down to the lowest temperature (1.6 K) of our experiment.
  This is different from the  As-system  Sr(Co$_{1-x}$Ni$_x$)$_2$As$_2$ where a non-Fermi liquid behavior has been observed near $x$ = 0.3 when the long-range magnetic order disappeared \cite{Sangeetha2019}. 
     Based on the analysis of the characteristic shape of the NMR spectra observed in the magnetically ordered state, we found that the ordered state is inhomogeneous where the magnitude of the ordered Co moments is modulated incommensurately with the lattice in Sr(Co$_{1-x}$Ni$_x$)$_2$P$_2$.   

 \section{Acknowledgments}
  We thank K. Rana for assistance in the initial stage of the NMR measurements. The research was supported by the U.S. Department of Energy, Office of Basic Energy Sciences, Division of Materials Sciences and Engineering. Ames National Laboratory is operated for the U.S. Department of Energy by Iowa State University under Contract No.~DE-AC02-07CH11358.


\begin{thebibliography}{99}
\bibitem{Korringa2} As the $x$ = 0.02 and 0.13 compounds exhibit the magnetically ordered states, the modified Korringa analysis may not be applicable to the compounds.   However, here we tentatively show the Korringa ratio to provide the idea of the evolution of the magnetic fluctuations.
\end{thebibliography}
\end{document}